\documentclass[pra,aps,10pt,twocolumn,showpacs,amsmath,amssymb]{revtex4-1}

\usepackage[english]{babel}
\usepackage{afterpage}
\usepackage{graphicx}% Include figure files
\usepackage{dcolumn}% Align table columns on decimal point
\usepackage{bm}% bold math
\usepackage{color}
\usepackage{slashed}%slashed D, k, A, etc.
\usepackage{latexsym}

\newcommand{\dd}{\mathrm{d}}         
\newcommand{\ii}{\mathrm{i}}  %complex i       
\newcommand{\ee}{\mathrm{e}}  %eksponent

\begin{document}

\title{Supercontinuum in ionization by relativistically intense and short laser pulses: ionization without interference and its time analysis}
\author{K. Krajewska$^{1,2}$}
\email[E-mail address:\;]{Katarzyna.Krajewska@fuw.edu.pl}
\author{J. Z. Kami\'nski$^1$}
\affiliation{$^1$Institute of Theoretical Physics, Faculty of Physics, University of Warsaw, Pasteura 5,
02-093 Warsaw, Poland\\ 
$^2$Department of Physics and Astronomy, University of Nebraska, Lincoln, Nebraska 68588-0299, USA
}
\date{\today}

\begin{abstract}
Ionization by relativistically intense laser pulses of finite duration is considered in the framework of strong-field quantum electrodynamics.
Our main focus is on the formation of ionization supercontinua. More specifically, when studying the energy distributions of photoelectrons ionized 
by circularly polarized pulses, we observe the appearance of broad structures lacking the interference patterns. These supercontinua extend over 
hundreds of driving photon energies, thus corresponding to high-order nonlinear processes. The corresponding polar-angle distributions show asymmetries
which are attributed to the radiation pressure experienced by photoelectrons. Moreover, our time analysis shows that the electrons comprising the supercontinuum
can form pulses of short duration. While we present the fully numerical results, their interpretation is based on the saddle-point approximation 
for the ionization probability amplitude.
\end{abstract}


\maketitle

\section{Introduction}
\label{sec::intro}

Interference of probability amplitudes is a fundamental quantum effect which, among others, manifests itself when a strong laser field interacts with matter. 
In his celebrated paper on strong-field ionization~\cite{Keldysh}, Keldysh has shown how the probability amplitudes emerging from the complex-time saddle points lead to the interference 
pattern in the angular-energy distribution of photoelectrons. This idea, further developed for instance in~\cite{Perelomov,Gribakin} (for related recent 
investigations, see~\cite{Arbo2010,Arbo2014,Arbo2014a}), has initiated theoretical investigations of such quantum processes like the above-threshold ionization, 
the rescattering phenomenon or the high-order harmonic generation. Along with important experimental achievements, it has led to a new branch of science called 
attosecond physics \cite{FT}, which has already found a lot of applications in physics, chemistry, biology, and medicine. Some of these achievements are described in the 
review articles (see, e.g.,~\cite{AgostiniMauro,Becker,KrauszIvanov,Maque,Dahl,Karnakov}) or in the recent collection of articles devoted to the Keldysh theory~\cite{KeldyshCel}.

The Keldysh approach has been further generalized in order to account for the interaction of ionized electrons with their parent ions. This has led to the concept 
of complex-time trajectories (i.e., trajectories along complex time determined by the purely classical Newton equations)~\cite{Popov1,Popov2,Popruzh}, or to the concept of quantum 
complex-time trajectories~\cite{CKK} which take into account the electron wave packet spreading or the quantum diffusion during the complex-time evolution, thus
eliminating problems with the rescattering trajectories present in the former approach. Theoretical methods that have been developed from the ideas put forward by Keldysh 
are usually called the Strong-Field Approximation (SFA) or the KFR theory (see, also~\cite{Faisal,Reiss}). Recently, with the development of lasers generating 
relativistically intense and very short pulses, similar interference patterns have been investigated in the relativistic strong-field quantum electrodynamics~\cite{FKK,PiazzaRev,Rosh}. 
In particular, such diverse phenomena have been studied as the vacuum polarization induced Young interference~\cite{King}, the Kapitza-Dirac effect~\cite{Muller,DM2015}, 
interference effects in Compton and Thomson scattering (see, e.g.,~\cite{KKscale,Wis2014}), and in laser-modified Mott scattering~\cite{ros1,ros2}, the coherent comb structures 
created by a finite train of short pulses in the Compton~\cite{KK1,KK2} and Breit-Wheeler~\cite{KK3} processes, which can be used for the diagnostics of relativistically intense laser pulses~\cite{KKdiag}.

On the other hand, the lack of interference may result in an appearance of supercontinuum~\cite{super1} which, since its demonstration in the early 
1970's, has been the focus of significant research activities. The supercontinuum generation has attracted a lot of attention owing to its enormous spectral broadening.
Thus, resulting in many useful applications, among others, in telecommunication and optical coherence tomography~\cite{super2}. It has been shown in Ref.~\cite{KK4} 
that the supercontinuum in the radiation domain spreading over keV or MeV energy regions can be generated during the Thomson or Compton scattering. In light of this result, the question arises: 
Is it possible to generate the supercontinuum in the ionization spectrum? In other words: Is it possible to choose the parameters of a driving pulse such 
that the energy spectrum of photoelectrons does not change rapidly with energy, on the scale of tens or even hundreds laser photon energies? 
In general, the answer to this question is negative as for even a few-cycle driving pulse the spectrum of photoelectrons
consists of a sequence of peaks separated approximately by the carrier laser frequency (which, for monochromatic plane waves, are called the multiphoton peaks). 
The Keldysh theory shows that such a structure arises as the result of interference of at least two complex-time probability amplitudes. Thus, in order to create 
the supercontinuum, one should have a dominant complex-time saddle point with an imaginary part which, over a broad range of electron final energies, 
would be much smaller than imaginary parts of the remaining saddle points. In other words, over a broad domain of electron final energies, interference 
of probability amplitudes should be suppressed. The aim of this paper is to show that, indeed, 
such a situation can happen in the relativistic ionization of atoms or positive ions by circularly polarized, short laser pulses.

The organization of this paper is as follows. In Sec.~\ref{sec:theory}, we present the theoretical formulation of relativistic ionization based on the Dirac equation 
and with the full account for the spin degrees of freedom. Specifically, in Sec.~\ref{sec:General} we develop a general approach, which for the velocity gauge is elaborated 
in detail in Sec.~\ref{sec:Velocity}. The supplementary Sec.~\ref{sec:Saddle} is devoted to the analysis of saddle points arising from our formulation, 
which further is used \textit{only} for the interpretation of numerical results. In Sec.~\ref{sec:Numerical}, we present numerical analysis of ionization of He$^+$ ions by short, 
circularly polarized and relativistically intense pulses defined in Sec.~\ref{sec:Pulse}. The energy and angular probability distributions are presented in Sec.~\ref{sec:Super}, 
where also the generation of ionization supercontinua is discussed. Sec.~\ref{sec:Azimuth} is devoted to the further analysis of the supercontinuum with the result 
that it can be shifted towards high energies by changing the azimuthal angle of emission. In Sec.~\ref{sec:global}, we study the energy dependence of the phase of probability 
amplitudes and show in Sec.~\ref{sec:spacetime} how the nearly linear dependence of the phase on the energy of emitted electrons leads to the time delay of the electron 
wave packets synthesized from a particular supercontinuum. In Sec.~\ref{sec:Conclusions}, we draw some concluding remarks. An Appendix~\ref{sec:units} contains 
supplementary materials concerning the normalization of the laser pulse shape functions and physical units.

Throughout the paper, we keep $\hbar=1$. Hence, the fine-structure constant equals $\alpha=e^2/(4\pi\varepsilon_0c)$. In numerical analysis we use relativistic units (rel. units) 
such that $\hbar=m_{\rm e}=c=1$ where $m_{\rm e}$ is the electron rest mass. We denote the product of any two four-vectors $a^{\mu}$ and $b^{\mu}$ as
$a\cdot b = a^{\mu}b_{\mu}=a^0b^0-a^1b^1-a^2b^2-a^3b^3$ ($\mu = 0,1,2,3$), where the Einstein summation convention is used. For the four-vectors we use both the contravariant 
$(a^0,a^1,a^2,a^3)$ and the standard $(a_0,a_x,a_y,a_z)=(a_0,\bm{a})$ notations. We employ the Feynman notation $\slashed{a} = \gamma\cdot a=\gamma^{\mu} a_{\mu}$ for the contraction 
with the Dirac matrices $\gamma^{\mu}$ and use a customary notation $\bar{u}=u^{\dagger}\gamma^0$, where $u^{\dagger}$ is the Hermitian conjugate of a bispinor $u$. 
Finally, we use the so-called light-cone variables. Namely, for a given space direction determined by a unit vector $\bm{n}$ (which in our paper is the direction of the laser 
pulse propagation) and for an arbitrary four-vector $a$, we keep the following notations: $a^{\|}=\bm{n}\cdot\bm{a}$, $a^-=a^0-a^{\|}$, $a^+=(a^0+a^{\|})/2$, and $\bm{a}^{\bot}=\bm{a}-a^{\|}\bm{n}$. 
Thus, $a\cdot b=a^+b^-+a^-b^+-\bm{a}^{\bot}\cdot\bm{b}^{\bot}$ and $\dd^4x=\dd x^+\dd x^-\dd^2x^{\bot}$.

\section{Theory and approximations}
\label{sec:theory}

\subsection{General theory}
\label{sec:General}

The relativistic ionization of one-electron atoms or ions is described by the Dirac equation ($e=-|e|$)
\begin{equation}
\bigl(\ii\gamma^{\nu}\partial_{\nu}-e\gamma^{\nu}\mathcal{A}_{\nu}(x)-m_{\mathrm{e}}c \bigr)\Psi(x)=0.
\label{dirac1}
\end{equation}
The electromagnetic potential $\mathcal{A}(x)$ is assumed to be of the form
\begin{equation}
\mathcal{A}^{\nu}(x)=\frac{1}{ec}V(\bm{x}){\delta^{\nu}}_0+A_{\mathrm{R}}^{\nu}(x),
\label{dirac2}
\end{equation}
where $V(\bm{x})$ is the binding potential and $A_{\mathrm{R}}^{\nu}(x)$ describes the laser pulse. The analysis of the time-evolution of the system leads to the exact expression for the probability amplitude
\begin{equation}
\mathcal{A}_{\mathrm{fi}}=-\ii \int\dd^4x \ee^{-\ii (E_0/c)x^0}\bar{\Psi}_{\mathrm{f}}(x)e\slashed{A}_{\mathrm{R}}(x)\Psi_{\mathrm{i}}(\bm{x}).
\label{dirac24}
\end{equation}
Here, the bispinor $\Psi_{\mathrm{i}}(\bm{x})$ describes the electron bound state of energy $E_0$ and $\Psi_{\mathrm{f}}(x)$ is the exact solution of Eq.~\eqref{dirac1} 
corresponding to the scattering state with the incoming spherical waves.

Although the analytic form of $\Psi_{\mathrm{i}}(\bm{x})$ is known for the Coulomb potential, the exact solution $\Psi_{\mathrm{f}}(x)$ can be determined only numerically. 
Recently, we observe a significant progress in solving numerically the relativistic Dirac equation (see, e.g., Refs.~\cite{Eva,ndirac1,ndirac2,ndirac3,press4}). Even so, for very large 
intensities of laser pulses available today (of the order of $10^{20}\mathrm{W/cm}^2$ and larger) and for high energies of photoelectrons (i.e., a few keV and higher) such solutions 
are not achievable. If, however, the exact scattering state is labeled by the asymptotic momentum $\bm{p}$ for which the kinetic energy, $\sqrt{(m_{\mathrm{e}}c^2)^2+(c\bm{p})^2}-m_{\mathrm{e}}c^2$, 
is much larger than the ionization potential of the initial bound state, $m_\mathrm{e}c^2-E_0$, then for the final scattering state the Born expansion with respect to the static potential $V(\bm{x})$ 
can be applied. This is the essence of the relativistic SFA which, in the lowest order, consists in replacing in Eq.~\eqref{dirac24} the exact solution $\Psi_{\mathrm{f}}(x)$ of Eq.~\eqref{dirac1} by the corresponding exact solution of the simplified Dirac equation
\begin{equation}
\bigl(\ii\slashed{\partial}-e\slashed{A}_{\mathrm{R}}(x)-m_{\mathrm{e}}c \bigr)\Psi^{(0)}_{\mathrm{f}}(x)=0.
\label{dirac25}
\end{equation}
For the relativistic SFA approach to ionization by plane wave fields, we refer the reader to~\cite{Sujata1,Sujata2,eikHeidelberg,Klaiber,eikHeidelberg2}. In this paper, however,
we focus on relativistic ionization by finite laser pulses. 

Let us choose as the solution of \eqref{dirac25}, denoted since now on by $\Psi^{(0)}_{\bm{p}\lambda}(x)$, the one with the well-defined electron momentum $\bm{p}$ and spin polarization 
$\lambda=\pm$. The probability amplitude $\mathcal{A}_{\mathrm{fi}}$~\eqref{dirac24}, which in this case we denote as $\mathcal{A}(\bm{p},\lambda;\lambda_{\mathrm{i}})$, becomes
\begin{align}
\mathcal{A}(\bm{p},\lambda;\lambda_{\mathrm{i}})=-\ii \int\frac{\dd^3q}{(2\pi)^3}\int\dd^4x\, & \ee^{-\ii q\cdot x}\bar{\Psi}^{(0)}_{\bm{p}\lambda}(x) \nonumber \\
 \times & e\slashed{A}_{\mathrm{R}}(x)\tilde{\Psi}_{\mathrm{i}}(\bm{q}),
\label{dirac26}
\end{align}
where
\begin{equation}
\Psi_{\mathrm{i}}(\bm{x})=\int\frac{\dd^3q}{(2\pi)^3}\ee^{\ii \bm{q}\cdot \bm{x}}\tilde{\Psi}_{\mathrm{i}}(\bm{q}).
\label{dirac26a}
\end{equation}
To shorten the notation, we have introduced $q=(q^0,\bm{q})=(E_0/c,\bm{q})$. Note that it is not the four-vector as it does not transform properly under the relativistic Lorentz transformations.
Nevertheless, for the sake of space, we shall call it the four-momentum as it is usually done for the electromagnetic potential $A^{\nu}_{\mathrm{R}}(x)$ in the Dirac equation \eqref{dirac1}, 
which is the four-vector only for particular gauges. In addition, $\lambda_{\mathrm{i}}$ labels the spin degrees of freedom for the initial state.

Even after this approximation, the direct numerical analysis of Eq.~\eqref{dirac26} is not possible as, for the laser pulses concentrated in a finite focus region, the numerical solution 
of Eq.~\eqref{dirac25} is not available for laser pulse and electron parameters mentioned above. For this reason, the so-called plane-wave front approximation for the laser beam is applied. 
It accounts for the finite time and space dependence of the laser pulse in the direction of its propagation, however, in the space directions perpendicular to the propagation direction it is assumed 
that the laser field extends to infinities. Such an approximation is justified if either the laser pulse is not tightly focused or interaction takes place with the highly energetic beams of 
particles, atoms or ions in the head-on kinematics (see, e.g., Ref.~\cite{Lee2010}). Note, that this approximation is commonly used in majority of investigations of quantum processes in strong 
laser fields, as only in this case it is possible to obtain the exact analytical solution of the Dirac equation, called the Volkov solution~\cite{Volkov}.

\subsection{Velocity gauge}
\label{sec:Velocity}

In nonrelativistic SFA one usually considers probability amplitudes in two gauges: the velocity and length gauges. The same is possible in the relativistic case. In this paper 
we consider the velocity gauge, postponing the consideration of the length gauge to the near future.

In order to derive the corresponding formulas let us consider the most general form (up to the gauge transformation) of the electromagnetic potential in the plane-wave front approximation,
\begin{equation}
A_{\mathrm{R}}(x)\equiv A(\phi)=A_0[\varepsilon_1 f_1(\phi)+\varepsilon_2 f_2(\phi)],
\label{dirac27}
\end{equation}
where $\phi=k\cdot x=k^0x^-$, $k=k^0n=k^0(1,\bm{n})$, $k^0=\omega/c$, $\omega=2\pi/T_{\mathrm{p}}$, $\varepsilon_j$ are two real polarization four-vectors normalized such that 
$\varepsilon_j\cdot\varepsilon_{j'}=-\delta_{jj'}$ and perpendicular to the propagation direction of the laser pulse, $k\cdot\varepsilon_j=0$. Here, we have also introduced 
$T_{\rm p}$ for the pulse duration. The two real functions, $f_j(\phi)$, called the shape functions, are arbitrary functions with the continuous second derivatives that vanish for 
$\phi<0$ and $\phi>2\pi$. Introducing the relativistically invariant parameter
\begin{equation}
\mu=\frac{|eA_0|}{m_{\mathrm{e}}c},
\label{dirac28}
\end{equation}
that defines the intensity of the laser field (the relation between this parameter and the time-averaged intensity of the laser pulse is presented in the Appendix~\ref{sec:units}),
we rewrite~\eqref{dirac27} as
\begin{equation}
eA(\phi)=-m_{\mathrm{e}}c\mu [\varepsilon_1 f_1(\phi)+\varepsilon_2 f_2(\phi)].
\label{dirac29}
\end{equation}
Hence, the Dirac equation \eqref{dirac25} becomes
\begin{equation}
\bigl\{\ii\slashed{\partial}+m_{\mathrm{e}}c\mu [\slashed{\varepsilon}_1 f_1(\phi)+\slashed{\varepsilon}_2 f_2(\phi)]-m_{\mathrm{e}}c \bigr\}\Psi(x)=0.
\label{dirac30}
\end{equation}
We denote as $\psi^{(+)}_{\bm{p}\lambda}(x)$ the Volkov solution of this equation for electrons (the superscript $(+)$ means that it is a positive-energy solution), where $\bm{p}$ is the electron asymptotic 
momentum and $\lambda=\pm$ labels the spin degrees of freedom. Its explicit form can be written as \cite{recoil}
\begin{align}
\psi^{(+)}_{\bm{p}\lambda}(x)=&\sqrt{\frac{m_{\mathrm{e}}c^2}{VE_{\bm{p}}}}\Bigl(1+\frac{m_{\mathrm{e}}c\mu}{2p\cdot k}\bigl[f_1(k\cdot x)\slashed{\varepsilon}_1\slashed{k} \nonumber \\
& + f_2(k\cdot x)\slashed{\varepsilon}_2\slashed{k} \bigr] \Bigr) \ee^{-\ii S_p^{(+)}(x)}u^{(+)}_{\bm{p}\lambda},
\label{dirac31}
\end{align}
where
\begin{align}
S_p^{(+)}(x)&= p\cdot x+\int_0^{k\cdot x}\dd\phi\Bigl[-\frac{m_{\mathrm{e}}c\mu}{p\cdot k}\bigl(\varepsilon_1\cdot p f_1(\phi) \nonumber \\
& + \varepsilon_2\cdot p f_2(\phi)\bigr)+\frac{(m_{\mathrm{e}}c\mu)^2}{2p\cdot k}\bigl(f_1^2(\phi)+f_2^2(\phi)\bigr)\Bigr].
\label{dirac32}
\end{align}
Here, the photoelectron asymptotic energy equals $E_{\bm{p}}=\sqrt{(c\bm{p})^2+(m_{\mathrm{e}}c^2)^2}$ whereas the on-mass-shell four-vector is $p=(p^0,\bm{p})=(E_{\bm{p}}/c,\bm{p})$. 
The Dirac free particle bispinors $u^{(+)}_{\bm{p}\lambda}$ are normalized such that $\bar{u}^{(+)}_{\bm{p}\lambda}u^{(+)}_{\bm{p}\lambda'}=\delta_{\lambda\lambda'}$, and $V$ is the quantization volume. 
With this normalization the final density of electron states, for a given spin degree of freedom, is equal to $V\dd^3p/(2\pi)^3$.

For our further purposes, we introduced the following functions (the explicit form of the ground state wave functions $\Psi_{\mathrm{i}}(\bm{x})$ for the hydrogenlike ions 
and for two spin polarizations can be found in the textbook~\cite{BjorkenDrell}): 
\begin{align}
B^{(0,0)}_{\bm{p}\lambda;\lambda_{\mathrm{i}}}(\bm{q})=&\bar{u}^{(+)}_{\bm{p}\lambda}\slashed{n}\tilde{\Psi}_{\mathrm{i}}(\bm{q}),
\nonumber \\
B^{(1,0)}_{\bm{p}\lambda;\lambda_{\mathrm{i}}}(\bm{q})=&\bar{u}^{(+)}_{\bm{p}\lambda}\slashed{\varepsilon}_1\tilde{\Psi}_{\mathrm{i}}(\bm{q}),
\nonumber \\
B^{(0,1)}_{\bm{p}\lambda;\lambda_{\mathrm{i}}}(\bm{q})=&\bar{u}^{(+)}_{\bm{p}\lambda}\slashed{\varepsilon}_2\tilde{\Psi}_{\mathrm{i}}(\bm{q}).
\label{dirac33}
\end{align}
This allows us to represent the probability amplitude \eqref{dirac26} in the form (note that $\slashed{\varepsilon}_j\slashed{k}\slashed{\varepsilon}_j=\slashed{k}$, for $j=1,2$, and $\slashed{\varepsilon}_1\slashed{k}\slashed{\varepsilon}_2+\slashed{\varepsilon}_2\slashed{k}\slashed{\varepsilon}_1=0$), 
\begin{align}
\mathcal{A}(\bm{p},\lambda;\lambda_{\mathrm{i}})=\int \frac{\dd^3q}{(2\pi)^3}\int\dd^4x \ee^{\ii S_p^{(+)}(x)-\ii q\cdot x} M(k\cdot x),
\label{dirac34}
\end{align}
with
\begin{align}
M(&k\cdot x)=\ii m_{\mathrm{e}}c\mu\sqrt{\frac{m_{\mathrm{e}}c^2}{VE_{\bm{p}}}} \nonumber \\
& \times \bigl[f_1(k\cdot x)B^{(1,0)}_{\bm{p}\lambda;\lambda_{\mathrm{i}}}(\bm{q})+f_2(k\cdot x)B^{(0,1)}_{\bm{p}\lambda;\lambda_{\mathrm{i}}}(\bm{q}) \nonumber \\
& -\frac{m_{\mathrm{e}}c\mu}{2p\cdot n}\bigl([f_1(k\cdot x)]^2+[f_2(k\cdot x)]^2\bigr)B^{(0,0)}_{\bm{p}\lambda;\lambda_{\mathrm{i}}}(\bm{q})\bigr].
\label{dirac34a}
\end{align}
Next, we introduce the so-called laser-dressed momentum \cite{KKC,KKBW}
\begin{align}
\bar{p}=p-&\frac{m_{\mathrm{e}}c\mu}{p\cdot k}(\varepsilon_1\cdot p\langle f_1\rangle+\varepsilon_2\cdot p\langle f_2\rangle )k \nonumber \\
+& \frac{(m_{\mathrm{e}}c\mu)^2}{2p\cdot k}(\langle f_1^2\rangle+\langle f_2^2\rangle )k,
\label{dirac35}
\end{align}
where, for any continuous function $F(\phi)$, that vanishes outside the interval $0\leqslant\phi\leqslant 2\pi$, we define
\begin{equation}
\langle F\rangle=\frac{1}{2\pi}\int_0^{2\pi}\dd\phi F(\phi).
\label{dirac36}
\end{equation}
The dressed momentum fulfills the equations
\begin{equation}
\bar{p}^-=p^- \quad \textrm{and}\quad \bar{\bm{p}}^{\bot}=\bm{p}^{\bot},
\label{dirac36a}
\end{equation}
that allow us to write
\begin{equation}
S_p^{(+)}(x)=\bar{p}^+x^-+p^-x^+-\bm{p}^{\bot}\cdot\bm{x}^{\bot}+G_p(k^0x^-),
\label{dirac37}
\end{equation}
where
\begin{align}
G_p(\phi)=&\int_0^{\phi}\dd\phi' \Bigl[-\frac{m_{\mathrm{e}}c\mu}{p\cdot k}\bigl(\varepsilon_1\cdot p (f_1(\phi')-\langle f_1\rangle) \nonumber \\
& + \varepsilon_2\cdot p (f_2(\phi')-\langle f_2\rangle)\bigr)+\frac{(m_{\mathrm{e}}c\mu)^2}{2p\cdot k}\bigl(f_1^2(\phi') \nonumber \\
& - \langle f_1^2\rangle+f_2^2(\phi')-\langle f_2^2\rangle\bigr)\Bigr].
\label{dirac38}
\end{align}
Inserting~\eqref{dirac37} into~\eqref{dirac34} we see that the integrations over $\dd x^+$ and $\dd^2x^{\bot}$ lead to the conservation conditions
\begin{equation}
p^-=q^- \quad \textrm{and}\quad \bm{p}^{\bot}=\bm{q}^{\bot},
\label{dirac38a}
\end{equation}
that permit us to carry out the integration over $\dd^3q$. The remaining integration over $\dd x^-$ is performed by applying the following Fourier decompositions for $j=1,2$ and $0\leqslant\phi=k^0x^-\leqslant 2\pi$,
\begin{align}
\bigl[f_1(\phi)\bigr]^j\exp[\ii G_p(\phi)]=&\sum_{N=-\infty}^{\infty}G_N^{(j,0)}\ee^{-\ii N\phi},\label{dirac39a} \\
\bigl[f_2(\phi)\bigr]^j\exp[\ii G_p(\phi)]=&\sum_{N=-\infty}^{\infty}G_N^{(0,j)}\ee^{-\ii N\phi} .\label{dirac39}
\end{align}
This allows us to write down the probability amplitude in the form
\begin{equation}
\mathcal{A}(\bm{p},\lambda;\lambda_{\mathrm{i}})=\ii m_{\mathrm{e}}c\mu\sqrt{\frac{m_{\mathrm{e}}c^2}{VE_{\bm{p}}}}\mathcal{D}(\bm{p},\lambda;\lambda_{\mathrm{i}}),
\label{dirac40}
\end{equation}
with
\begin{align}
\mathcal{D}(\bm{p},\lambda;\lambda_{\mathrm{i}})&=\sum_{N=-\infty}^{\infty}\frac{\ee^{2\pi\ii(\bar{p}^+-q^+-Nk^0)/k^0}-1}{\ii (\bar{p}^+-q^+-Nk^0)} \nonumber \\
&\times \Bigl[G^{(1,0)}_N B^{(1,0)}_{\bm{p}\lambda;\lambda_{\mathrm{i}}}(\bm{Q})
+ G^{(0,1)}_N B^{(0,1)}_{\bm{p}\lambda;\lambda_{\mathrm{i}}}(\bm{Q})  \nonumber \\
&-\frac{m_{\mathrm{e}}c\mu}{2p\cdot n}[G^{(2,0)}_N+G^{(0,2)}_N]B^{(0,0)}_{\bm{p}\lambda;\lambda_{\mathrm{i}}}(\bm{Q})\Bigr]
\label{dirac41}
\end{align}
and 
\begin{equation}
\bm{Q}=\bm{p}+(q^0-p^0)\bm{n}.
\label{dirac42}
\end{equation}
Note that in the corresponding nonrelativistic SFA and with the dipole approximation applied to the laser field the momentum 
$\bm{Q}$ in Eq.~\eqref{dirac42} is not shifted by the vector $(q^0-p^0)\bm{n}$, independently of the gauge choice. This term, among others, is responsible for the so-called radiation pressure~\cite{press0} which 
has recently been discussed in the literature (see, e.g.,~\cite{press1,press2,press3,Chelkowski,press4}). Such a momentum shift appears also in the so-called Coulomb-Corrected relativistic SFA (see, e.g., Eq.~(24) in 
Ref.~\cite{eikHeidelberg2}).

Eq.~\eqref{dirac41} allows us to define the most probable energy absorbed by photoelectrons during the ionization. Namely, the ionization probability distribution is maximum for such $N$ for which 
the denominator, $\bar{p}^+-q^+-Nk^0$, takes the smallest values. This happens for those $N$ which are as close as possible to the number
\begin{equation}
N_{\mathrm{eff}}=\frac{c\bar{p}^+-cq^+}{\omega}=N_{\mathrm{osc}}\frac{c\bar{p}^+-cq^+}{\omega_{\mathrm{L}}}.
\label{dirac42a}
\end{equation}
Hence, the energy which is most probably transferred from the pulse to the atomic or ionic system during the ionization equals
\begin{equation}
E_{\mathrm{tr}}=N_{\mathrm{eff}}\omega=\frac{N_{\mathrm{eff}}}{N_{\mathrm{osc}}}\omega_{\mathrm{L}}.
\label{dirac42b}
\end{equation}
The number $N_{\mathrm{eff}}/N_{\mathrm{osc}}$ estimates how many laser photons, each carrying the energy $\omega_{\mathrm{L}}$, have to be absorbed for the photoelectron to be detected with 
the final energy $E_{\bm{p}}$. In numerical illustrations presented in this paper these numbers range from 2000 up to 3000, which shows the high degree of nonlinearity of processes being
considered. Note that both quantities $N_{\mathrm{eff}}$ and $E_{\mathrm{tr}}$ are gauge-invariant~\cite{KKBW}. Moreover, $N_{\mathrm{eff}}$ is invariant for relativistically covariant processes.

Finally, the total probability of ionization equals
\begin{equation}
P(\lambda;\lambda_{\mathrm{i}})=\mu^2\frac{(m_{\mathrm{e}}c)^3}{(2\pi)^3}\int\frac{\dd^3p}{p^0}|\mathcal{D}(\bm{p},\lambda;\lambda_{\mathrm{i}})|^2,
\label{dirac43}
\end{equation}
whereas its triply-differential distribution takes the form
\begin{equation}
\frac{\dd^3P(\bm{p},\lambda;\lambda_{\mathrm{i}})}{\dd E_{\bm{p}}\dd^2\Omega_{\bm{p}}}=\mu^2\frac{(m_{\mathrm{e}}c)^3}{(2\pi)^3c}|\bm{p}|\cdot|\mathcal{D}(\bm{p},\lambda;\lambda_{\mathrm{i}})|^2.
\label{dirac44}
\end{equation}
We see that in the last two formulas the quantization volume $V$ cancels. For the purpose of numerical illustrations let us introduce the dimensionless distribution,
\begin{equation}
\mathcal{P}_{\lambda_{\mathrm{i}}\lambda}(\bm{p})=\alpha^2m_{\mathrm{e}}c^2\frac{\dd^3P(\bm{p},\lambda;\lambda_{\mathrm{i}})}{\dd E_{\bm{p}}\dd^2\Omega_{\bm{p}}},
\label{dirac45}
\end{equation}
which is the probability distribution in the atomic units.
Later on, we shall label the spin degrees of freedom as $\downarrow$ and $\uparrow$ for $\lambda$ or $\lambda_{\mathrm{i}}$ equal to $-$ and $+$, respectively.

In this Section, we have derived formulas for the energy-angular probability distributions that are valid \textit{only} for sufficiently 
large electron kinetic energies $E_{\mathrm{kin}}$, for which the condition:
\begin{equation}
E_{\mathrm{kin}}=\sqrt{(m_{\mathrm{e}}c^2)^2+(c\bm{p})^2}-m_{\mathrm{e}}c^2 \gg m_{\mathrm{e}}c^2-E_0
\label{dirac46}
\end{equation}
is satisfied. Thus, we shall apply current theory to ionization of He$^+$ ions, with the ionization potential 
of roughly 54eV, and we will analyze the energy-angular probability distributions for final electron kinetic energies larger than 1keV.
It is also commonly assumed that the SFA is applicable for sufficiently intense laser fields. Specifically, when the ponderomotive energy is larger 
or comparable to the ionization potential. This condition is also very well-fulfilled in our numerical analysis.

\subsection{Saddle-point analysis}
\label{sec:Saddle}

Expressions for the energy-angular probability distribution of photoelectrons, which have been derived in the previous Section, are not convenient for the interpretation of calculated results. 
On the other hand, a very appealing interpretation can be provided by analyzing the saddle points of the corresponding integrands, as it has been suggested by Keldysh.
The latter has been applied, for instance, in Refs.~\cite{Arbo2010,CKK} in investigations of the diffraction/interference structures in the probability distributions. 
As the formulas presented above do not suit for such an analysis, therefore, we have to rewrite the expression for the probability amplitude $\mathcal{A}(\bm{p},\lambda;\lambda_{\mathrm{i}})$ 
in a different form. For this purpose we present \eqref{dirac34} in the light-cone variables as
\begin{align}
\mathcal{A}(\bm{p},\lambda;\lambda_{\mathrm{i}})=&\frac{1}{k^0}\int_0^{2\pi}\dd\phi \int\frac{\dd^3q}{(2\pi)^3}\int\dd x^+\dd^2x^{\bot} \nonumber \\
&\times \ee^{\ii (p^--q^-)x^+ -\ii(\bm{p}^{\bot}-\bm{q}^{\bot})\cdot\bm{x}^{\bot}} \nonumber \\
&\times\ee^{\ii G(g_0,g_1,g_2,h;\phi)}M(\phi),
\label{saddle1}
\end{align}
where $M(\phi)$ is defined by \eqref{dirac34a} and
\begin{align}
G(g_0,g_1,g_2,h;\phi)=\int_0^{\phi}\dd\phi' & \bigl[g_0+g_1f_1(\phi')+g_2f_2(\phi')   \nonumber \\
 & +h\bigl(f_1^2(\phi')+f_2^2(\phi')\bigr)\bigr],
\label{saddle2}
\end{align}
with
\begin{align}
g_0&=\frac{p^+-q^+}{k^0}, \nonumber \\
g_j&=-m_{\mathrm{e}}c\mu\frac{\varepsilon_j\cdot p}{k\cdot p}, \quad j=1,2, \nonumber \\
h&=\frac{(m_{\mathrm{e}}c\mu)^2}{2k\cdot p}.
\label{saddle3}
\end{align}
Since the integrations over $\dd x^+\dd^2x^{\bot}$ lead to the conservation conditions \eqref{dirac38a}, we end up with the following expression for the probability amplitude,
\begin{equation}
\mathcal{A}(\bm{p},\lambda;\lambda_{\mathrm{i}})=\frac{1}{k^0}\int_0^{2\pi}\dd\phi\,\ee^{\ii G(g_0,g_1,g_2,h;\phi)}\bigl[M(\phi)\bigr]_{\bm{q}=\bm{Q}},
\label{saddle4}
\end{equation}
where $\bm{Q}$ is defined in Eq.~\eqref{dirac42}. This is the formula which suits for the saddle-point analysis. In the following, if it does not lead to misunderstandings, 
we abbreviate the function $G(g_0,g_1,g_2,h;\phi)$ by $G(\phi)$. Let us also note that, due to the conservation conditions~\eqref{dirac38a}, $g_0$ in~\eqref{saddle3} equals
\begin{equation}
g_0=\frac{p^0-q^0}{k^0},
\label{saddle5}
\end{equation}
and it depends only on energies of the initial and final states.

Applying now the standard asymptotic procedure for the approximate evaluation of integrals, we determine the saddle points by solving the equation
\begin{equation}
G'(\phi)=0,
\label{saddle6}
\end{equation}
where `\textit{prime}' means the derivative over $\phi$. This equation has in general complex solutions. Among them we select only those saddle points, denoted by $\phi_s$, for which 
$\textrm{Im}\,G(\phi_s)>0$. Hence, we arrive at the approximate expression for the probability amplitude,
\begin{equation}
\mathcal{A}(\bm{p},\lambda;\lambda_{\mathrm{i}})=\frac{1}{k^0}\sum_s \ee^{\ii G(\phi_s)}\sqrt{\frac{2\pi\ii}{G^{\prime\prime}(\phi_s)}}\bigl[M(\phi_s)\bigr]_{\bm{q}=\bm{Q}}.
\label{saddle7}
\end{equation}
If $\bigl[M(\phi)\bigr]_{\bm{q}=\bm{Q}}$ is singular at the saddle point, we have to apply the so-called singular saddle-point approximation described, for instance, in Refs.~\cite{Gribakin,Popruzh,CKK}. 
However, independently of the method applied, the dominant behavior of the integral~\eqref{saddle4} is determined by the exponent $\ee^{\ii G(\phi_s)}$ which usually, for high energy electrons, 
decays very fast to 0. Moreover, the analysis of such integrals shows that in most cases at least two saddle points contribute significantly to the above sum, which results in the interference pattern 
observed in the probability distributions. At this point, let us emphasize that it is not our aim to compare the results predicted by the exact formula~\eqref{dirac44} with the ones that follow from 
the saddle-point approximation~\eqref{saddle7}. We treat the expression~\eqref{saddle7} only as the appealing interpretative tool for our numerical analysis.

Eq.~\eqref{saddle7} suggests that the interference pattern in ionization is suppressed (i.e., the ionization supercontinuum may appear) if there is only one saddle point for which, over a broad range
of electron energies, $\textrm{Im}\,G(\phi_s)$ is much smaller than the corresponding values for the remaining saddle points. We shall see below that such a situation 
can indeed take place. Note that conclusions which follow from this interpretation are gauge-independent, as only the function $M(\phi)$ depends on the chosen gauge.

\section{Numerical analysis}
\label{sec:Numerical}

\subsection{Pulse shape}
\label{sec:Pulse}

The laser pulse with the $\sin^2$ envelope considered in this paper is define as follows. First, we introduce two angles,
\begin{equation}
\delta_j=(j-1)\frac{\pi}{2}, \quad j=1,2,
\label{shapes1}
\end{equation}
and two functions,
\begin{equation}
F_j(\phi)=F_0(\phi,\delta_j,\chi)\cos(\delta+\delta_j),
\label{shapes2}
\end{equation}
with
\begin{equation}
F_0(\phi,\delta_j,\chi)=N_0\sin^2\Bigl(\frac{\phi}{2}\Bigr)\sin(N_{\mathrm{osc}}\phi+\delta_j+\chi)
\label{shapes3}
\end{equation}
for $0<\phi <2\pi$ and 0 otherwise. Next, we define the shape functions $f_j(\phi)$ of the electromagnetic vector potential~\eqref{dirac29} as
\begin{equation}
f_j(\phi)=-\int_0^{\phi}\dd \phi' F_j(\phi').
\label{shapes4}
\end{equation}
Since for $N_{\mathrm{osc}}>1$ the Fourier decompositions of $F_j(\phi)$ do not contain constant terms, the above definitions of $f_j(\phi)$ guarantee that also for $\phi > 2\pi$ the vector potential vanishes. 
The physical interpretation of the remaining parameters is as follows. The angle $\delta$ determines the polarization properties of the laser pulse and, for the circularly polarized field, we choose 
$\delta=\pi/4$. The carrier envelope phase $\chi$ is assumed to be $\pi/2$. Finally, $N_0$ is the normalization-dependent real and positive factor which is chosen such that the normalization 
condition~\eqref{un2} is fulfilled. The integer $N_{\mathrm{osc}}$ defines the number of cycles in the pulse and it is assumed to be equal to 4. 
The laser pulse propagates in the $z$ direction ($\bm{k}=(\omega/c)\bm{n}$, $\bm{n}=\bm{e}_z$) and the real polarization vectors are equal to $\bm{\varepsilon}_1=\bm{e}_x$ and $\bm{\varepsilon}_2=\bm{e}_y$.

\begin{figure}
\includegraphics[width=5cm]{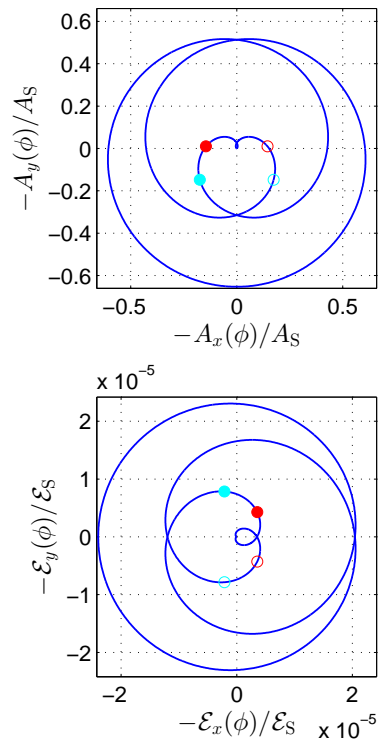}
\caption{(Color online) Trajectories of the tips of the electromagnetic vector potential $\bm{A}(\phi)$ (upper panel) and the electric field vector $\bm{\mathcal{E}}(\phi)$ (lower panel) 
in the relativistic units for the laser pulse parameters discussed in detail in Sec.~\ref{sec:Pulse}. All trajectories start from and end up at the origin $(0,0)$. In order to show 
the direction of the time-evolution we mark with the filled and open circles the points on these trajectories corresponding to some particular values of $\phi$: red (dark) 
filled circle for $\phi=0.32\pi$, cyan (gray) filled circle for $\phi=0.396\pi$, cyan (gray) open circle for $\phi=2\pi-0.396\pi$, and red (dark) open circle for $\phi=2\pi-0.32\pi$. 
The meaning of these particular points is discussed in the text. We observe the azimuthal symmetry of the electromagnetic potential, $\varphi\rightarrow \pi-\varphi\textrm{ mod }2\pi$ or $(x,y)\rightarrow (-x,y)$, 
and of the electric field, $\varphi\rightarrow -\varphi\textrm{ mod }2\pi$ or $(x,y)\rightarrow (x,-y)$.
\label{shapesbisss20150531}}
\end{figure}

We consider here the interaction of a laser pulse with a positively charged He$^+$ ion, which is a one-electron ion with $Z=2$. Since in our case the pulse is finite, therefore, the initial 
bound state is well-defined. For this reason, we do not have to restrict our analysis to one-electron ions of very large $Z$, as it has been done in Refs.~\cite{eikHeidelberg,Klaiber,eikHeidelberg2} 
where the infinite plane-wave field has been considered. Moreover, as the ion can have relativistic energy, we can choose the carrier frequency of the laser pulse, $\omega_{\mathrm{L}}$, freely. 
This is due to the fact that the calculations are carried out in the ion reference frame in which, for the head-on geometry of the ion and the laser beam, both the frequency and the electric-field strength 
are Doppler-upshifted. This aspect has been originally discussed for the laser-induced Bethe-Heitler process in Ref.~\cite{CartenBH} (see, also the review articles~\cite{FKK,PiazzaRev}). For this reason, we assume 
that $\omega_{\mathrm{L}}=N_{\mathrm{osc}}\omega=20$eV and the time-averaged intensity is $I=10^{20}$~W/cm$^2$ (as it has been chosen for instance in Ref.~\cite{Sujata1}); in other words, 
we consider parameters which are experimentally available~\cite{xray1,xray2,xray3}.

In Fig.~\ref{shapesbisss20150531} we present trajectories of the tips of vectors $\bm{A}(\phi)$ (upper panel) and $\bm{\mathcal{E}}(\phi)=-\partial_t\bm{A}(\phi)$ (lower panel) in the polarization 
plane $xy$, which is perpendicular to the direction of propagation of the laser pulse $\bm{n}$. The plots are in the relativistic units, with the units for the electromagnetic potential 
$A_{\mathrm{S}}=m_{\mathrm{e}}c/|e|$ and the electric field strength $\mathcal{E}_{\mathrm{S}}=m_{\mathrm{e}}c^2/(|e|\lambdabar_{\mathrm{C}})=m_{\mathrm{e}}^2c^3/|e|$; the latter is known as 
the Sauter-Schwinger critical field~\cite{Sauter,Schwinger}. Hence, in Fig.~\ref{shapesbisss20150531}, we present $x$ and $y$ components of the vectors 
$-\bm{A}(\phi)/A_\mathrm{S}=e\bm{A}(\phi)/(m_{\mathrm{e}}c)$ and $-\bm{\mathcal{E}}(\phi)/\mathcal{E}_{\mathrm{S}}=e\bm{\mathcal{E}}(\phi)/(|e|\mathcal{E}_{\mathrm{S}})$.

The intensity of the laser pulse is considered to be relativistic if the time-averaged ponderomotive energy (see, e.g., \cite{KKdiag}),
\begin{equation}
U=-e^2\frac{\langle A\cdot A\rangle - \langle A\rangle\cdot\langle A\rangle}{2m_{\mathrm{e}}},
\label{shapes5}
\end{equation}
is comparable to or larger than $m_{\mathrm{e}}c^2$, or the amplitude of the electromagnetic potential is comparable to or larger than $A_{\mathrm{S}}$. As follows from the upper panel
of Fig.~\ref{shapesbisss20150531}, for the chosen laser pulse parameters the laser pulse intensity can be considered as nearly relativistic.

\subsection{Supercontinuum}
\label{sec:Super}
\begin{figure}
\includegraphics[width=6cm]{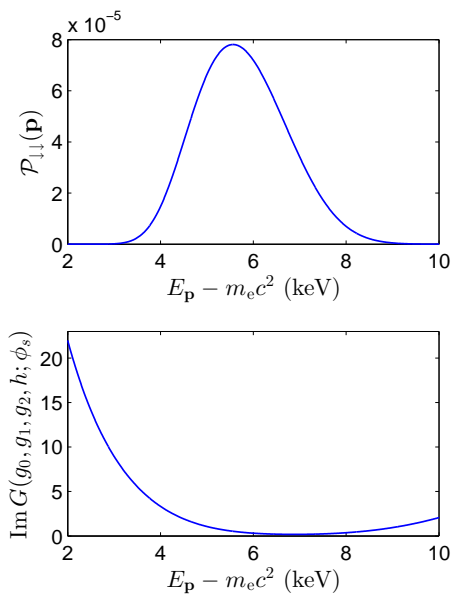}
\caption{(Color online) The ionization probability $\mathcal{P}_{\downarrow\downarrow}(\bm{p})$ (upper panel) of He$^+$ ions for $\theta_{\bm{p}}=0.48\pi$ and $\varphi_{\bm{p}}=0$ as a function 
of the photoelectron kinetic energy $E_{\bm{p}}-m_{\mathrm{e}}c^2$, for the laser pulse parameters described in Sec.~\ref{sec:Pulse}. We observe a very broad supercontinuum correlated 
with the minimum of the imaginary part of $G(g_0,g_1,g_2,h;\phi_s)$ for a particular saddle point (lower panel). ${\rm Im}\,G(g_0,g_1,g_2,h;\phi_s)$ for the remaining saddle points are 
at least two orders of magnitude larger.
\label{ereiss1d20150523}}
\end{figure}
\begin{figure}
\includegraphics[width=6cm]{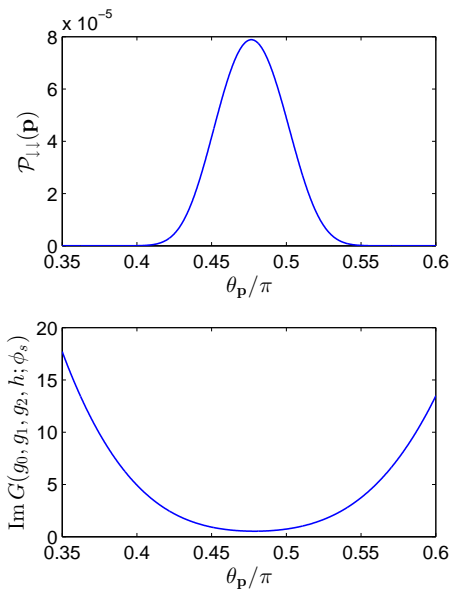}
\caption{(Color online) Upper panel shows the ionization probability $\mathcal{P}_{\downarrow\downarrow}(\bm{p})$ as a function of $\theta_{\bm{p}}$ for He$^+$ ions with energy $E_{\bm{p}}-m_{\mathrm{e}}c^2=5565$eV 
and $\varphi_{\bm{p}}=0$, and for the laser pulse parameters described in Sec.~\ref{sec:Pulse}. As the result of the radiation pressure, the distribution is shifted towards smaller $\theta_{\bm{p}}$ angles 
compared to the expectations based on the nonrelativistic theory. This result agrees with the saddle-point analysis of the probability amplitude (lower panel), where we see that the maximum of 
the probability amplitude coincides with the minimum of ${\rm Im}\,G(g_0,g_1,g_2,h;\phi_s)$.
\label{treiss1d20150523}}
\end{figure}

Previously, the main reason to investigate ionization of atoms or ions within the relativistic quantum mechanics was to analyze the electron spin effects, not present in its nonrelativistic counterpart. 
It has been shown in Ref.~\cite{Sujata1} that the spin-flipping processes, in which the electron initial and final spins both projected on the same direction in space (usually chosen as the direction 
of propagation of the laser pulse) are opposite to each other, are less probable by roughly two orders of magnitude (i.e., of the order of the fine structure constant $\alpha$) as compared to 
the ionization processes in which the projections of the initial and final electron spins are conserved. Our numerical analysis confirms these findings also for very short pulses. Therefore, 
in the remaining part of this paper we shall analyze the process in which the initial and final spins are anti-parallel to the propagation direction of the laser pulse. The second dominant 
process, with the spin projections parallel to the laser field propagation direction, only marginally differs from the first one.

Recently, due to the experimental results reported in Ref.~\cite{press1} concerning the effects related to the radiation pressure exposed on the atomic or ionic systems by intense laser pulses, 
we observe an increased interest in theoretical investigations of ionization with the relativistic effects accounted for~\cite{press4,press2,press3,Chelkowski}. We shall discuss below that the signatures 
related to the radiation pressure are present in the developed above theoretical approach [cf. our comments below Eq.~\eqref{dirac42}], although the main focus of our studies is on the formation 
of a broad ionization supercontinuum.

We consider the probability distribution of electrons, $\mathcal{P}_{\downarrow\downarrow}(\bm{p})$, in a given space direction defined by the polar and azimuthal angles, 
$\theta_{\bm{p}}=0.48\pi$ and $\varphi_{\bm{p}}=0$, respectively. For photoelectron kinetic energies a little bit smaller that 1keV, which still might be in the domain of applicability of theoretical 
methods developed above, we observe a typical oscillatory dependence of $\mathcal{P}_{\downarrow\downarrow}(\bm{p})$ on the energy. This can be interpreted as the interference of probability 
amplitudes emerging from at least two saddle points. These oscillations, however, gradually disappear with increasing the photoelectron energy. Such that, in the broad part of the spectrum covering few hundreds of the laser carrier frequencies $\omega_{\mathrm{L}}$, 
we observe a smooth behavior of the probability distribution with the clearly visible maximum for the electron kinetic energy $E_{\bm{p}}-m_{\mathrm{e}}c^2\approx 5565$eV, as presented in 
Fig.~\ref{ereiss1d20150523} (upper panel). As anticipated above, the interpretation of these findings can be based on the analysis of saddle points of the probability amplitude~\eqref{saddle4}. 
Indeed, in the whole domain, extending from 3keV up to 9keV, there is only one saddle point of imaginary part of $G(g_0,g_1,g_2,h;\phi_s)$ very close to 0, as compared to the remaining saddle 
points for which ${\rm Im}\,G(g_0,g_1,g_2,h;\phi_s)$ are at least two orders of magnitude larger.

Let us analyze further the angular distribution for a particular kinetic energy $E_{\bm{p}}-m_{\mathrm{e}}c^2=5565$eV, that approximately corresponds to the maximum in the supercontinuum 
presented in Fig.~\ref{ereiss1d20150523}. First, we consider the polar angle distribution for $\varphi_{\bm{p}}=0$. The nonrelativistic theory predicts that for circularly polarized laser field 
ionization occurs predominantly in the plane perpendicular to the direction of propagation. In our case the maximum of the probability distribution should appear for $\theta_{\bm{p}}=\pi/2$. 
However, for sufficiently intense laser pulses, due to the so-called radiation pressure, the maximum should be shifted towards the direction of propagation, i.e., towards smaller angles $\theta_{\bm{p}}$. 
This effect is indeed visible as presented in Fig.~\ref{treiss1d20150523}, and can be also attributed to the minimum of ${\rm Im}\,G(g_0,g_1,g_2,h;\phi_s)$.

\begin{figure}
\includegraphics[width=6cm]{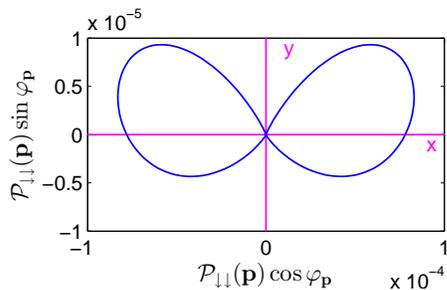}
\caption{(Color online) The ionization probability $\mathcal{P}_{\downarrow\downarrow}(\bm{p})$ of He$^+$ ions for $E_{\bm{p}}-m_{\mathrm{e}}c^2=5565$eV and $\theta_{\bm{p}}=0.48\pi$ 
as the function of $\varphi_{\bm{p}}$, for the laser pulse parameters described in Sec.~\ref{sec:Pulse}. The vertical scale is much smaller than the horizontal one, meaning 
that the distribution is tightly elongated around the $x$ direction.
}
\label{preiss1d20150524}
\end{figure}

\begin{figure}
\includegraphics[width=6cm]{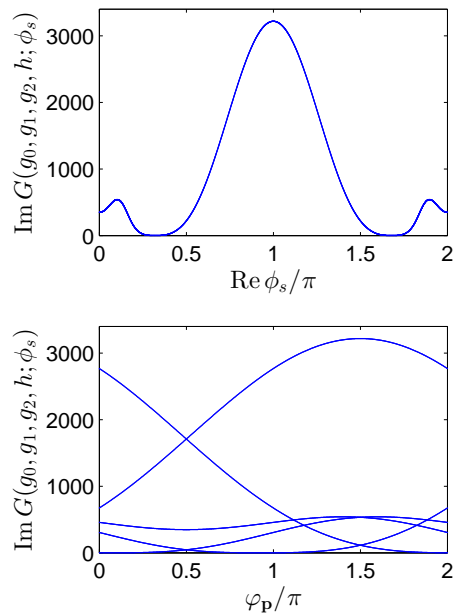}
\caption{(Color online) Shows ${\rm Im}\,G(g_0,g_1,g_2,h;\phi_s)$ for $E_{\bm{p}}-m_{\mathrm{e}}c^2=5565$eV and $\theta_{\bm{p}}=0.48\pi$. 
We observe that, for a given azimuthal angle $\varphi_{\bm{p}}$, there are five saddle points (lower panel). 
Among them only one, let us say $\phi_1$, has a very small positive imaginary part of $G(g_0,g_1,g_2,h;\phi_s)$ which is observed for $\varphi_{\bm{p}}=0.066\pi$ and $\varphi_{\bm{p}}=\pi-0.066\pi$. 
These values correspond to the real parts of the saddle points $\textrm{Re}\,\phi_1=1.68\pi$ and $\textrm{Re}\,\phi_1=0.32\pi$, respectively, as shown in the upper panel.
}
\label{saddlepoints0bisspr1d20150531}
\end{figure}

Although the polar angle distribution of ionization has an expected shape if we account for the laser radiation pressure, the azimuthal angle distribution for relativistic intensities 
and the high-energy part of the spectrum differs from the predictions of the nonrelativistic SFA. In the nonrelativistic theory for moderately intense laser fields, the ionization occurs with the largest 
probability when the electric field strength of the laser pulse is maximum (see, e.g., Ref.~\cite{CKK}). Hence, Fig.~\ref{shapesbisss20150531} would suggest that the azimuthal angle distribution should 
be nearly isotropic. Such pattern is not observed in the relativistic SFA, as presented in Fig.~\ref{preiss1d20150524}. For this particular electron final energy the azimuthal distribution is peaked 
for $\varphi_{\bm{p}}=0.066\pi$ and $\varphi_{\bm{p}}=\pi-0.066\pi$. One can interpret this result by performing the saddle-point analysis, presented in Fig.~\ref{saddlepoints0bisspr1d20150531}. 
In the lower panel we observe that, for each of the above azimuthal angles, there is one saddle point [we denote it as $\phi_1(\varphi_{\bm{p}})$] for which the imaginary part of 
$G(g_0,g_1,g_2,h;\phi_1)$ is very small. It follows from the upper panel of Fig.~\ref{saddlepoints0bisspr1d20150531} that these particular azimuthal angles and the real parts of the corresponding 
saddle points can be grouped in pairs: $(\varphi_{\bm{p}},{\rm Re}\,\phi_1(\varphi_{\bm{p}}))=(0.066\pi,1.68\pi)$ and $(\pi-0.066\pi,0.32\pi)$. It is usually assumed that the real parts of the saddle points 
determine the escape time of electrons from atoms or ions. The values of the electromagnetic vector potential and the electric field strength for these two particular phases ${\rm Re}\,\phi_1(\varphi_{\bm{p}})$ 
are marked in Fig.~\ref{shapesbisss20150531} by red (dark) open and filled circles. The positions of these points show that the escape of electrons for relativistic intensities takes place not when the electric 
field strength is maximum (as it is in the nonrelativistic SFA), but for times when the pulse ramps on and off. This happens at least for ions with not very large $Z$.

\subsection{Azimuthal angle dependence}
\label{sec:Azimuth}

The width and the position of maximum of the ionization supercontinuum can be controlled by changing the azimuthal angle. It appears that, for the considered laser pulse shape and the azimuthal angle 
from the upper part of the $(x,y)$ plane, i.e., for $0<\varphi_{\bm{p}}<\pi$, the supercontinuum can be shifted towards lower energies. As a result, the interference pattern starts building up. If, however, 
we choose $\varphi_{\bm{p}}$ from the interval $(\pi,2\pi)$, the smooth supercontinuum appears for larger energies. These findings can be also explain by applying the saddle-point analysis.

\begin{figure}
\includegraphics[width=6cm]{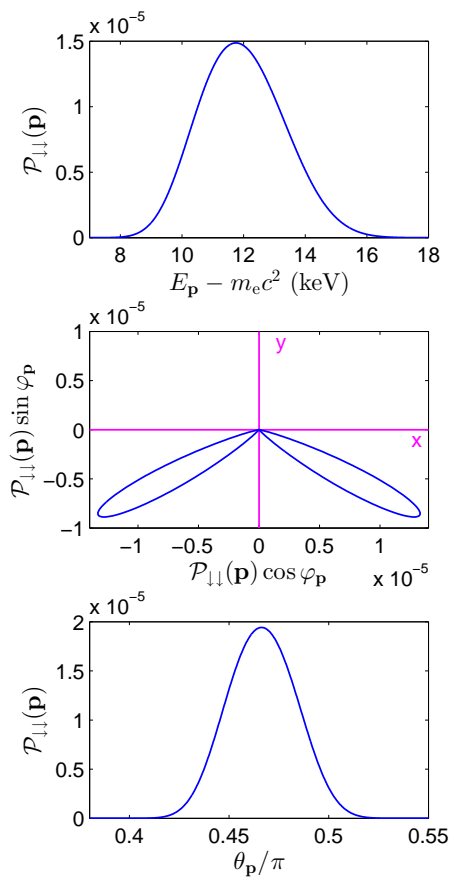}
\caption{(Color online) The same as in Figs.~\ref{ereiss1d20150523},~\ref{preiss1d20150524} and~\ref{treiss1d20150523}, but respectively for:
$\theta_{\bm{p}}=0.48\pi$ and $\varphi_{\bm{p}}=1.8\pi$ (top panel), $E_{\bm{p}}-m_{\mathrm{e}}c^2=11750$eV and $\theta_{\bm{p}}=0.48\pi$ (middle panel), and 
$E_{\bm{p}}-m_{\mathrm{e}}c^2=11750$eV and $\varphi_{\bm{p}}=1.8\pi$ (bottom panel).
}
\label{eptreiss2d20150524}
\end{figure}

\begin{figure}
\includegraphics[width=6cm]{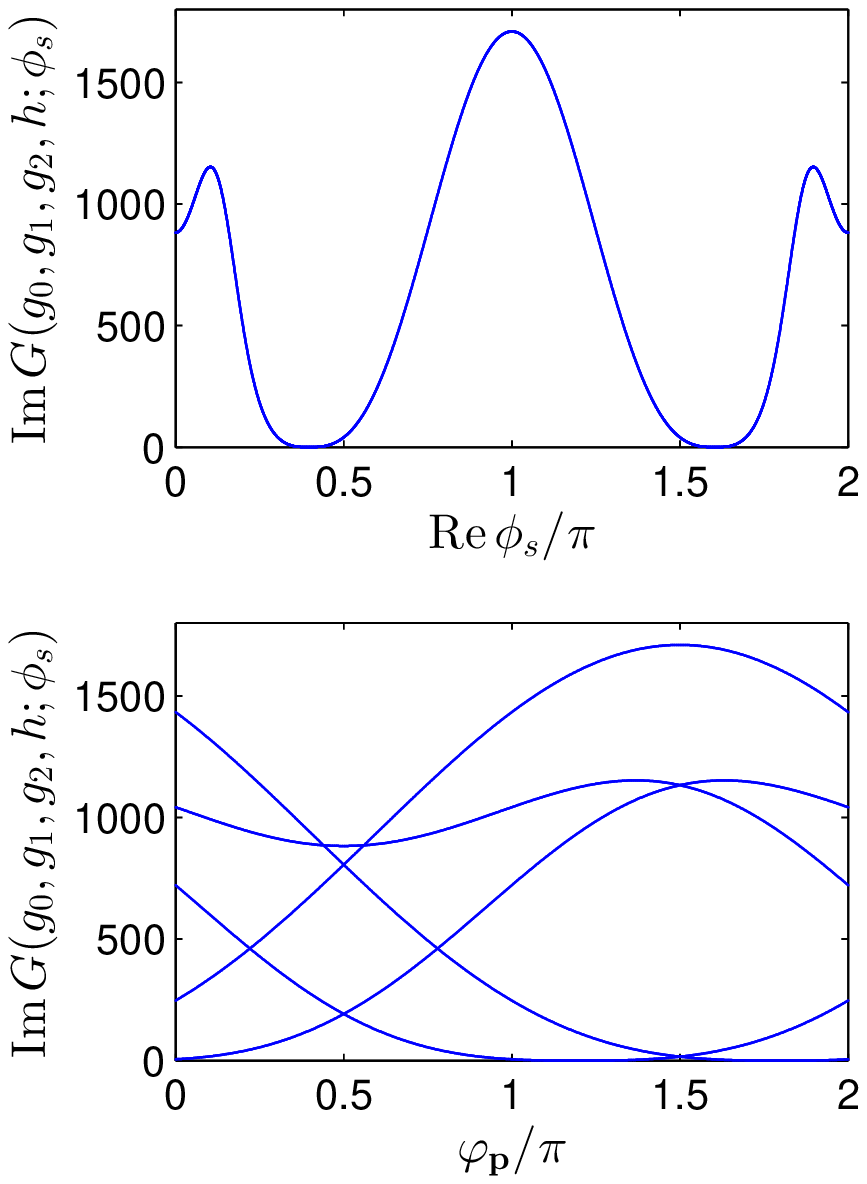}
\caption{(Color online) The same as in Fig.~\ref{saddlepoints0bisspr1d20150531}, but for $E_{\bm{p}}-m_{\mathrm{e}}c^2=11750$eV and $\theta_{\bm{p}}=0.48\pi$.
}
\label{saddlepoints2bisspr1d20150531}
\end{figure}

In order to illustrate this effect let us consider the energy spectrum of photoelectrons ejected in the direction given by spherical angles $\theta_{\bm{p}}=0.48\pi$ and $\varphi_{\bm{p}}=1.8\pi$.  
This spectrum is presented in the top panel of Fig.~\ref{eptreiss2d20150524}. We see that now the supercontinuum extends from 8keV up to 16keV with the maximum for $E_{\bm{p}}-m_{\mathrm{e}}c^2\approx 11750$eV. 
By fixing now the photoelectron kinetic energy, $E_{\bm{p}}-m_{\mathrm{e}}c^2= 11750$eV and the polar angle, $\theta_{\bm{p}}=0.48\pi$, in the middle panel of Fig.~\ref{eptreiss2d20150524} 
we plot the azimuthal angle distribution of ionization probability. As we see, the sidelobes have been turned downwards, but still the symmetry $x\rightarrow -x$ is preserved. As in the previous case, 
we also see the effect of radiation pressure as the maximum of the polar angle distribution is shifted to the left (the bottom panel of Fig.~\ref{eptreiss2d20150524}).

In Fig.~\ref{saddlepoints2bisspr1d20150531}, we present a similar analysis of the imaginary part of $G(g_0,g_1,g_2,h;\phi_s)$ as the function of ${\rm Re}\,\phi_s$ (upper panel) and 
$\varphi_{\bm{p}}$ (lower panel). The minima of this function are for $(\varphi_{\bm{p}},{\rm Re}\,\phi_1(\varphi_{\bm{p}}))=(1.2\pi,0.396\pi)$ and $(1.8\pi,1.604\pi)$, that correspond to cyan (gray) filled and open circles in Fig.~\ref{shapesbisss20150531}.

\subsection{Global phase of ionization amplitude}
\label{sec:global}

\begin{figure}
\includegraphics[width=7cm]{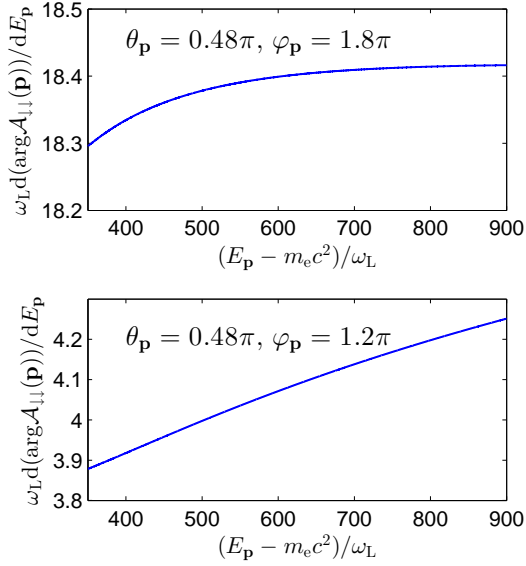}
\caption{(Color online) The derivative of the phase $\Phi_{\mathcal{A}}$ calculated over the electron energy, Eq.~\eqref{phase1}, for two directions of emission
denoted in each panel. For these angles, the energy distributions are identical and are presented in the upper panel of Fig.~\ref{eptreiss2d20150524}.
}
\label{ereiss2phasedir20150628}
\end{figure}

The energy-angular distribution of photoelectrons discussed above depends only on the modulus squared of the probability amplitude, $|\mathcal{A}(\bm{p},\lambda;\lambda_{\mathrm{i}})|^2$. However, the probability amplitude, as a complex function, is also characterized by the phase, $\Phi_{\mathcal{A}}(\bm{p},\lambda;\lambda_{\mathrm{i}})$,
\begin{align}
\mathcal{A}(\bm{p},\lambda;\lambda_{\mathrm{i}})&=\exp\bigl[\ii\Phi_{\mathcal{A}}(\bm{p},\lambda;\lambda_{\mathrm{i}})\bigr]|\mathcal{A}(\bm{p},\lambda;\lambda_{\mathrm{i}})|, \label{phase1a} \\
\Phi_{\mathcal{A}}(\bm{p},\lambda;\lambda_{\mathrm{i}})&=\mathrm{arg}\bigl[\mathcal{A}(\bm{p},\lambda;\lambda_{\mathrm{i}})\bigr],
\label{phase1}
\end{align}
defined up to a constant term not affecting any physically observable quantities. In particular, this constant term disappears in the derivative of the phase $\Phi_{\mathcal{A}}$ over the electron energy. 
This derivative is presented in Fig.~\ref{ereiss2phasedir20150628} for two directions of the electron momentum $\bm{p}$, for which the azimuthal angular distribution shown in the middle panel of 
Fig.~\ref{eptreiss2d20150524} is maximum, i.e., for the angles $(\theta_{\bm{p}},\varphi_{\bm{p}})$ equal to $(0.48\pi,1.8\pi)$ and $(0.48\pi,1.2\pi)$. We see that these derivatives are nearly constant,
over a very broad range of electron kinetic energies. This means that the phase $\Phi_{\mathcal{A}}$ can be well-represented for this supercontinuum energy region by the linear dependence,
\begin{equation}
\Phi_{\mathcal{A}}(\bm{p},\lambda;\lambda_{\mathrm{i}})\approx
\Phi_0(\bm{n}_{\bm{p}},\lambda;\lambda_{\mathrm{i}})+\Phi_1(\bm{n}_{\bm{p}},\lambda;\lambda_{\mathrm{i}})E_{\bm{p}},
\label{phase2}
\end{equation}
where $\bm{n}_{\bm{p}}=\bm{p}/|\bm{p}|$.

In order to elucidate what are the consequences of this approximate formula, we mention the effects originating from the frequency-dependent phase, similar to~\eqref{phase2}, in classical and quantum optics.
It is known that the space- and time-dependence of radiation pulses follow from their frequency distributions (see, e.g.,~\cite{KK4,frog}). 
For radiation pulses, the constant term $\Phi_0$ does not influence the angular-energy and space-time distributions. On contrary, the linear term leads to the time-delay 
of pulses; i.e., the larger $\Phi_1$ the larger time-delay is observed. This suggests that, as long as Fig.~\ref{ereiss2phasedir20150628} is concerned, the electron wave packet released 
during ionization in the direction $(\theta_{\bm{p}},\varphi_{\bm{p}})=(0.48\pi,1.8\pi)$ is delayed with respect to the one propagating in the direction 
$(\theta_{\bm{p}},\varphi_{\bm{p}})=(0.48\pi,1.2\pi)$. This is what the above saddle-point analysis predicts, and we are going to address this problem in the next Section.

\subsection{Space and time analysis}
\label{sec:spacetime}

The probability amplitude for the electron to be found at the point $x$ equals
\begin{equation}
\Psi[x,\lambda;\lambda_{\mathrm{i}}|\mathcal{F}]=\int\frac{V\dd^3p}{(2\pi)^3}\sqrt{\frac{m_{\mathrm{e}}c^2}{VE_{\bm{p}}}}\ee^{-\ii p\cdot x}u^{(+)}_{\bm{p}\lambda}\mathcal{A}(\bm{p},\lambda;\lambda_{\mathrm{i}})\mathcal{F}(\bm{p}).
\label{spacetime1}
\end{equation}
Here, it is assumed that the probability amplitude is calculated for such space-time points $x$, for which the action of the laser pulse is over, i.e., $A(k\cdot x)=0$. 
The so-called filter function $\mathcal{F}(\bm{p})$ selects only such final electron momenta, from which we are going to build up the wave packet. For our purpose, 
we select only those electrons that move in a given space direction $\bm{n}_0$ and have energies from a chosen range. To be more specific, we assume that
\begin{align}
\mathcal{F}(\bm{p})&=\theta(E_{\mathrm{max}}-E_{\bm{p}}+m_{\mathrm{e}}c^2) \nonumber \\
&\times\theta(E_{\bm{p}}-m_{\mathrm{e}}c^2-E_{\mathrm{min}})\delta^{(2)}(\Omega_{\bm{p}}-\Omega_{\bm{n}_0}),
\label{spacetime2}
\end{align}
where $\theta(E)$ is the Heaviside step function, $E_{\mathrm{min}}=7\mathrm{keV}$ and $E_{\mathrm{max}}=18\mathrm{keV}$ [cf. the top panel of Fig.~\ref{eptreiss2d20150524}]. 
Hence, the electron wave packet propagating in the space direction $\bm{x}=d\bm{n}_0$ (where $d$ is the distance from the nucleus), and with initial and final spins anti-parallel 
to the laser pulse propagation direction, is proportional to
\begin{align}
\mathcal{A}_{\downarrow}(t,d)\sim \int_{E_{\mathrm{min}}+m_{\mathrm{e}}c^2}^{E_{\mathrm{max}}+m_{\mathrm{e}}c^2}&\dd E_{\bm{p}}\, \ee^{-\ii E_{\bm{p}}t+\ii |\bm{p}|d}u^{(+)}_{|\bm{p}|\bm{n}_0,\downarrow} \nonumber \\
\times & \sqrt{E_{\bm{p}}}\, |\bm{p}|\mathcal{A}(|\bm{p}|\bm{n}_0,\downarrow;\downarrow),
\label{spacetime3}
\end{align}
where we have neglected all irrelevant prefactors. Finally, the time-distance probability distribution of these selected electrons is equal to
\begin{equation}
\mathcal{P}_{\downarrow}(t,d)=\bigl[\mathcal{A}_{\downarrow}(t,d)\bigr]^{\dagger}\mathcal{A}_{\downarrow}(t,d).
\label{spacetime4}
\end{equation}

\begin{figure}
\includegraphics[width=7cm]{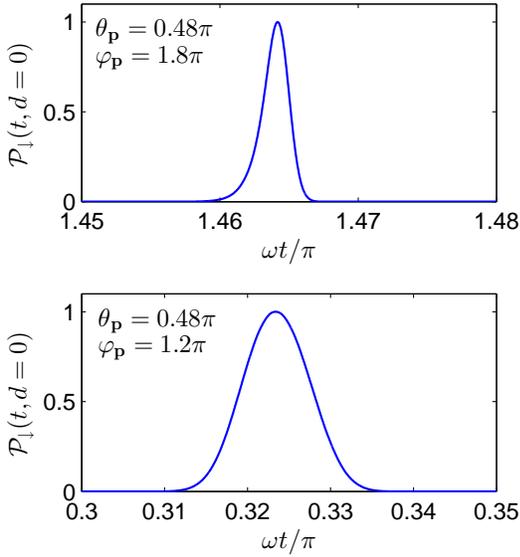}
\caption{(Color online) The normalized probability distribution~\eqref{spacetime4} for electrons escaping in the directions defined by the polar and azimuthal angles as denoted in each panel,
and for the remaining parameters same as in Figs.~\ref{eptreiss2d20150524} and~\ref{ereiss2phasedir20150628}. 
}
\label{ereiss2spec20150629}
\end{figure}

Although the last definition applies to positive times $t$ and distances $d$ such that $k\cdot x=\omega t-(\omega/c)d\bm{n}\cdot\bm{n}_0>2\pi$, nevertheless, by extrapolating its validity to $d=0$ 
(where the nucleus is located) we can roughly estimate the escape time from the ion of the selected electrons, as it is presented in Fig.~\ref{ereiss2spec20150629}. We can judge from the positions 
of maxima that the escape times for the selected group of photoelectrons are around $\omega t=1.464\pi$ for the direction $(\theta_{\bm{p}},\varphi_{\bm{p}})=(0.48\pi,1.8\pi)$ and $\omega t=0.323\pi$ 
for the direction $(\theta_{\bm{p}},\varphi_{\bm{p}})=(0.48\pi,1.2\pi)$. This agrees reasonably well with the predictions based on the saddle-point analysis, where we have estimated these times to be 
$\omega t=\mathrm{Re}\,\phi_1=1.604\pi$ and $\omega t=\mathrm{Re}\,\phi_1=0.396\pi$, respectively.

\begin{figure}
\includegraphics[width=7cm]{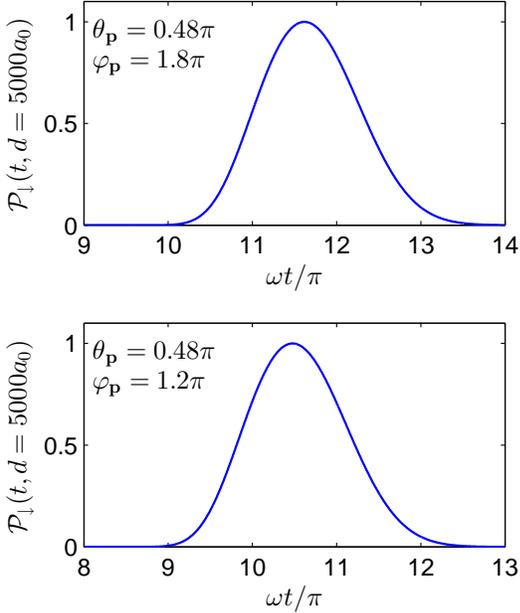}
\caption{(Color online) The same as in Fig.~\ref{ereiss2spec20150629} but for the distance $d=5000a_0$ from the nucleus, where $a_0$ is the Bohr radius. 
}
\label{ereiss2spec20150629a}
\end{figure}

The width of the distributions presented in Fig.~\ref{ereiss2spec20150629} are not larger than $10^{-2}$. This means that, at the time of escape, 
electrons from these supercontinua are created in very short pulses lasting for
\begin{equation}
\Delta t\approx 10^{-2}\pi\frac{1}{\omega}=\frac{10^{-2}\pi}{5\mathrm{eV}/\hbar}\approx 10^{-17}\mathrm{s}.
\label{spacetime5}
\end{equation}
Then, during the time-evolution, the electron wave packets spread. However, even at the distance $5000a_0$ from the ion their time durations are of the order of $10^{-15}\mathrm{s}$, as it is shown in 
Fig.~\ref{ereiss2spec20150629a}. As expected, the pulses presented there are delayed with respect to each other by the time approximately equal to $1.15\pi/\omega$. 
This quite well agrees with predictions of the Keldysh theory.

\section{Conclusions}
\label{sec:Conclusions}

We have developed the theoretical formulation for relativistic ionization driven by an arbitrary laser pulse. This has been done
within the lowest order Born approximation with respect to the atomic potential. As an illustration, we have studied ionization of ${\rm He}^+$ 
by a short, relativistically intense, circularly polarized pulse. As our analysis shows, it is possible to adjust parameters of
the pulse and the target system such that high-order nonlinear processes result in appearance of a broad supercontinuum in the energy spectrum of photoelectrons.

We have further analyzed the properties of supercontinuum electrons. We have checked that they are predominantly ionized through the spin-conserved
processes. We have also observed that, in contrast to the nonrelativistic ionization by circularly polarized pulses, the polar-angle distributions of photoelectrons are
asymmetric. This has been ascribed to the radiation pressure experienced by electrons. As we have argued, in the energy region of supercontinuum,
the total phase of the probability amplitude of ionization can be approximated as a linear function of the photoelectron energy. Therefore, the electron
pulses composed out of this part of the spectrum can be delayed with respect to each other. More importantly, the supercontinuum electrons can form very short 
pulses. Despite the fact that they are spreading in time, we have demonstrated that these pulses remain fairly short.

We have shown numerically that relativistic ionization can lead to the formation of electron supercontinua. This is particularly
interesting in the context of designing new sources of electron pulses. While our analysis is based on purely numerical treatment, we have
also performed an analysis of probability amplitudes based on approximation via the saddle-point method. The latter has shown, for instance, that
relativistic ionization can occur with significant probabilities at the pedestal of the driving pulse. Not, like in nonrelativistic ionization, at the
pulse maximum. Other predictions of the saddle-point approximation have agreed well with our fully numerical results.

\section*{Acknowledgements}

This work is supported by the Polish National Science Center (NCN) under Grant No. 2012/05/B/ST2/02547. 
K.K. also acknowledges the support from the Kosciuszko Foundation. We would like to thank Prof. Sujata Bhattacharyya for discussions.

\appendix*
\section{Remarks on physical units}
\label{sec:units}

The time-averaged intensity of a pulse described in the plane-wave front approximation equals
\begin{equation}
I=(\langle {f'_1}^2\rangle+\langle {f'_2}^2\rangle)\varepsilon_0 c\frac{(\omega m_{\mathrm{e}}c)^2}{e^2}\mu^2,
\label{un1}
\end{equation}
where $\omega$ is related to the pulse duration $T_{\rm p}$ such that $T_{\mathrm{p}}=2\pi/\omega$. If the number of cycles is $N_{\mathrm{osc}}$
then the carrier frequency is $\omega_{\mathrm{L}}=N_{\mathrm{osc}}\omega$, and in our analysis we keep it fixed. In order to have the time-averaged 
intensity independent of the number of cycles in the pulse we choose the normalization of the shape functions such that
\begin{equation}
\langle {f'_1}^2\rangle+\langle {f'_2}^2\rangle=\frac{1}{2}N_{\mathrm{osc}}^2 
\label{un2}
\end{equation}
and, consequently,
\begin{equation}
I=\frac{1}{2}\varepsilon_0 c\frac{(\omega_{\mathrm{L}} m_{\mathrm{e}}c)^2}{e^2}\mu^2.
\label{un3}
\end{equation}
We can rewrite this relation as
\begin{equation}
I=I_{\mathrm{rel}}\Bigl(\frac{\hbar\omega_{\mathrm{L}}}{m_{\mathrm{e}}c^2}\Bigr)^2\mu^2,
\label{un4}
\end{equation}
where we have restored the Planck constant $\hbar$, and where $I_{\mathrm{rel}}$ is the relativistic unit of intensity,
\begin{equation}
I_{\mathrm{rel}}=\frac{m_{\mathrm{e}}c^3}{8\pi\alpha\lambdabar_C^3}=\frac{m^4_{\mathrm{e}}c^6}{8\pi\alpha\hbar^3}\approx 2.324\times 10^{29}\mathrm{W/cm}^2.
\label{un5}
\end{equation}
Here, $\lambdabar_C=\hbar/m_{\mathrm{e}}c$ is the Compton wavelength divided by $2\pi$. On the other hand, the nonrelativistic unit of intensity is equal to
\begin{equation}
I_{\mathrm{nrel}}=\alpha^6 I_{\mathrm{rel}}\approx 3.51\times 10^{16}\mathrm{W/cm}^2.
\label{un6}
\end{equation}
Thus,
\begin{equation}
I=I_{\mathrm{nrel}}\Bigl(\frac{\hbar\omega_{\mathrm{L}}}{\alpha^2 m_{\mathrm{e}}c^2}\Bigr)^2\mu^2_{\mathrm{nrel}},
\label{un7}
\end{equation}
where $\alpha^2 m_{\mathrm{e}}c^2\approx 27.21\mathrm{eV}$ is the nonrelativistic unit of energy and
\begin{equation}
\mu_{\mathrm{nrel}}=\frac{\mu}{\alpha}=\frac{|eA_0|}{\alpha m_{\mathrm{e}}c}.
\label{un8}
\end{equation}
Above, $\alpha m_{\mathrm{e}}c=\hbar/a_0$ is the nonrelativistic unit of momentum and $a_0$ is the Bohr radius.

Eqs.~\eqref{un4} and~\eqref{un7} allow one to relate the parameter $\mu$ in~\eqref{dirac28} to the time-averaged 
intensity of the laser pulse given in the standard units of $\mathrm{W/cm}^2$.

\end{document}